\documentclass[conference]{IEEEtran}
\IEEEoverridecommandlockouts

\usepackage{cite}
\usepackage{amsmath,amssymb,amsfonts}
\usepackage{algorithmic}
\usepackage{graphicx}

\usepackage{textcomp}
\usepackage{amsmath,amssymb,amsfonts}
\usepackage{algorithmic,algorithm}
\usepackage{graphicx}
\usepackage{tabularx}
\usepackage{subcaption}
\usepackage{amsmath,amsfonts}
\usepackage{textcomp}
\usepackage{array}
\usepackage{url}
\usepackage{stfloats}
\usepackage{textcomp}
\usepackage{ragged2e, mdframed}
\usepackage{cite}
\usepackage{amsmath,amssymb,amsfonts}
\usepackage{algorithmic}
\usepackage{graphicx}
\usepackage{textcomp}
\usepackage{xcolor}
\usepackage{caption}
\usepackage{multirow}
\usepackage{verbatim}
\usepackage{multirow}
\usepackage{xcolor}
\def\BibTeX{{\rm B\kern-.05em{\sc i\kern-.025em b}\kern-.08em
    T\kern-.1667em\lower.7ex\hbox{E}\kern-.125emX}}

\begin{document}

\title{SPEEDNet: Salient Pyramidal
Enhancement Encoder-Decoder Network for Colonoscopy Images
}

\author{\IEEEauthorblockN{1\textsuperscript{st} Given Name Surname}
\IEEEauthorblockA{\textit{dept. name of organization (of Aff.)} \\
% City, Country
}
\and
\IEEEauthorblockN{2\textsuperscript{nd} Given Name Surname}
\IEEEauthorblockA{\textit{dept. name of organization (of Aff.)} \\
% City, Country
}
\and
\IEEEauthorblockN{3\textsuperscript{rd} Given Name Surname}
\IEEEauthorblockA{\textit{dept. name of organization (of Aff.)} \\
% City, Country
}

}
\author{\IEEEauthorblockN{Tushir Sahu$^1$, Vidhi Bhatt$^2$, Sai Chandra Teja R$^3$, Sparsh Mittal$^4$, Nagesh Kumar S$^5$}
\IEEEauthorblockA{
\textit{$^1$ Indian Institute of Information Technology (IIIT) Jabalpur, $^2$ Gujarat Technological University, $^3$Independent Researcher}\\ 
\textit{$^4$Indian Institute of Technology (IIT) Roorkee, $^5$SVIMS Tirupati}\\
\{tushirsahu22,vidhibhatt3008,saichandrateja\}@gmail.com,sparsh.mittal@ece.iitr.ac.in,nageshkumarsingaram@gmail.com}

\thanks{While serving as interns at IIT Roorkee, Tushir and Vidhi made contributions to this study.}
 }

    \maketitle

%within medical images, stand as pivotal goals in the realm of 
%\emph{Approach:} 

\begin{abstract}
Accurate identification and precise delineation of regions of significance, such as tumors or lesions, is a pivotal goal in medical imaging analysis. This paper proposes SPEEDNet, a novel architecture for precisely segmenting lesions within colonoscopy images. SPEEDNet uses a novel block named ``Dilated-Involutional Pyramidal Convolution Fusion'' (DIPC). A DIPC block combines the dilated involution layers pairwise into a pyramidal structure to convert the feature maps into a compact space.  
This lowers the total number of parameters while improving the learning of representations across an optimal receptive field, thereby reducing the blurring effect. 
On the EBHISeg dataset, SPEEDNet outperforms three previous networks: UNet, FeedNet, and AttesResDUNet. Specifically, SPEEDNet attains an average dice score of 0.952 and a recall of 0.971.
Qualitative results and ablation studies provide additional insights into the effectiveness of SPEEDNet. The model size of SPEEDNet is 9.81 MB, significantly smaller than that of UNet (22.84 MB), FeedNet (185.58 MB), and AttesResDUNet (140.09 MB).
\end{abstract}

\begin{IEEEkeywords}
Artificial intelligence (AI) for medical diagnosis, deep neural network, encoder-decoder network, Dilated-Involution.
\end{IEEEkeywords}

\section{Introduction}
Colon cancer is a significant global health concern. It is the second most prominent contributor to cancer-related fatalities worldwide and the third most prevalent malignancy in both men and women.
For its detection and diagnosis, both non-invasive screening and invasive diagnostic procedures have been used. 
Accurate colon image segmentation is vital for precise cancer detection.
A deep learning-based computer vision method holds promise for refined pathological diagnosis and prognosis. This has motivated researchers to propose several deep-learning techniques. 

%new
Detection of colon cancer isn't as simple and straightforward as  it seems. A few of the key hurdles include:  Asymptotic early stages, Screening barriers due to financial constraints, age-related risk, location variability, small polyps, overlapping symptoms, fear and stigma, and many more. One of the major challenges with colon cancer is that it often begins with no symptoms in its early stages. Due to this, colon cancer is frequently found in its late stages, when few effective treatments are available and the prognosis is dismal. Therefore, the creation of accurate and effective technologies for colon cancer early detection is urgently needed.
Early detection through regular screenings can significantly improve the chances of successful treatment and improved outcomes for colon cancer patients.

Histopathological diagnosis is vital for categorizing colon tissue as normal or abnormal, blending histopathology expertise with AI-driven analysis. This approach eases the pathologist's workload and enhances diagnostic efficiency. Deep learning techniques hold promise for refined pathological diagnosis and prognosis \cite{litjens2016deep}.

For colon image segmentation, Tajbakhsh et al. \cite{tajbakhsh2015automated} introduce a context-based information-based system for accurately locating only colonic polyps without considering other categories. SegNet \cite{badrinarayanan2017segnet} employs pooling layers. However,  they may compromise spatial resolution that is crucial for extracting tiny features such as complicated colon cancer areas. Jha et al. \cite{jha2019resunet++} achieve notable improvements in colorectal polyp segmentation, particularly for smaller image sets. Graham et al. \cite{graham2019mild} suggest a dilated network with low information loss. However, it is ineffective in distinguishing between histological components that are extremely similar. Dumitru et al. \cite{dumitru2023using} present a feature-rich design but do not address the class imbalance and complexity issue. Overall, several issues still need to be solved, such as imprecise borders, lower predictive performance, sensitivity to image quality, and high model size.

%While affecting individuals of all ages, it is more commonly observed among older adults. The disease originates from genetic mutations in benign cells of the large intestine, triggering uncontrolled proliferation and the formation of tumors. 
%Various methods are used for the detection and diagnosis of this at different stages. These techniques range from non-invasive screening methods to more invasive diagnostic procedures. 

%Prior research works \cite{jha2019resunet++,venugopal2022dtp,dumitru2023using} advanced colon segmentation, yet challenges like imprecise borders, class imbalance, sensitivity to image quality and computational complexity persist. Addressing these issues is key to dependable colon image segmentation enhancing diagnostics and patient care. 
\begin{figure*}[htbp]
  \centering  \includegraphics[width=1.0\textwidth,height=7.7 cm]{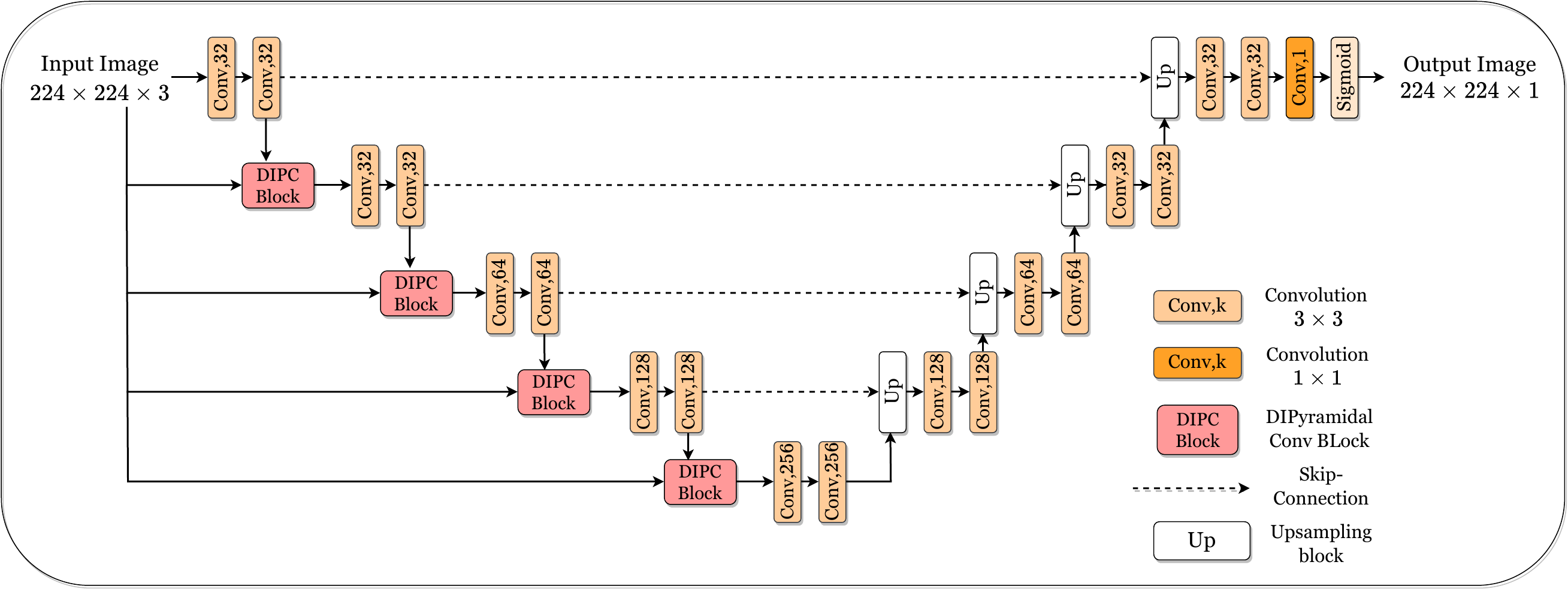}   
  \captionsetup{font=sf} 
  \caption{Overall architecture of SPEEDNet, k shows the number of filters.}
 \label{fig:speednet}
\end{figure*}
In this paper, we propose SPEEDNet, a novel architecture for precisely segmenting lesions within colonoscopy images. By incorporating dilated involution at different pyramid levels, the model adapts its receptive field to different object scales within the image. This ensures that the network is proficient in capturing the nuanced edges of objects, regardless of whether they are fine and intricate or large and prominent.

We summarize our contributions as follows:

1. SPEEDNet incorporates a DIPC Block. It merges dilated involution and convolution components to enhance segmentation by capturing contextual details and refining object features.

2. SPEEDNet's predictive performance has been rigorously evaluated on the EBHI-Seg dataset \cite{shi2023ebhi} using dice coefficient, Jaccard index, precision, and recall. SPEEDNet outperforms three previous networks (UNet, FeedNet, and AttResDUNet) on nearly all classes and metrics. For example, it attains an average dice score of 0.952 and a recall of 0.971.

3. SPEEDNet has a model size of 9.81 MB, whereas UNet, FeedNet, and AttesResDUNet have model sizes of 22.84 MB, 185.58 MB, and 140.09 MB, respectively. Clearly, SPEEDNet provides high segmentation efficiency with only small resource utilization.

\section{Proposed Method \& Architecture}
\textbf{Overall architecture:} Fig. \ref{fig:speednet} shows the architecture of SPEEDNet, which draws inspiration from UNet \cite{ronneberger2015u}.
 SPEEDNet includes a five-level encoder with a Dilated-Involutional Pyramidal Convolution Fusion (DIPC) block and two convolution blocks. The decoder also has five layers from bottom to top, each with a pair of  $ 3 \times 3$ convolution blocks and an upsampling block. 
 After each convolution, RELU activation and post-activation BatchNorm layers are applied. The critical difference between UNet and SPEEDNet is that SPEEDNet includes a mechanism for attention-driven feature enhancement. This mechanism takes feature maps from different scales, along with downsampled versions of the input image and computes attention maps.

A previous work, FeedNet \cite{deshmukh2022feednet}, uses LSTM layers to capture temporal relationships inside fixed-size context windows. However, incorporating an LSTM with a context window that encompasses the entire image results in a model size of 185.58 MB. 
By contrast, SPEEDNet incorporates a DIPC block, which is augmented with dilated involutions. 
Involution operation is spatial-specific and channel-agnostic, whereas convolutions are spatial-agnostic and channel-specific. 
By virtue of generating spatially-adapted kernels, involution operation effectively minimizes the channel redundancies that are commonly encountered in convolutions. The spatially focused nature of involution leads to less number of parameters while maintaining or even improving performance.  
Hence, involution offers a compelling advantage over networks such as UNet and its variants that employ traditional convolution. This has led to increasing adoption of involution in recent years, particularly in the design of lightweight architectures.

\textbf{DIPC Block:} The DIPC block uses varied dilation rates to enhance the receptive field for specific involution and convolution layers in a pyramidal manner.
This helps in efficiently capturing diverse image patterns and intricate details. It decreases the amount of blurring in the semantic segmentation map by merging local and global salient features, which are then aligned via downsampling. Notably, previous networks, viz., UNet, FeedNet, and AttesResDU-Net, fail to reduce blurring as effectively as SPEEDNet.
The DIPC block combines saliency maps with varied dilation rates using element-wise pair summing, which helps retain multi-scale information across the network. 
Element-wise summation produces a complete representation of salient features.

The use of dilated convolutions helps in extraction of significant characteristics. The multiplication of pool maps and the attention map introduces dynamic information flow. This allows the network to flexibly route information based on the feature saliency, which enhances segmentation.
 Vakanski et al. \cite{vakanski2020attention} employ convolution-based attention modules, especially with large kernels that lead to spatial information loss. Another limitation of their work is that its effectiveness is contingent upon the quality of salient feature maps generated from the input image. Using low-quality maps can degrade predictive performance. The proposed DIPC Block puts emphasis on integrating the salient maps and the feature maps from previous stage of the encoder with feed-forward connection. Thus, the DIPC block generates efficient salient maps capturing important feature representation.

Consider a $224 \times 224 \times c$ feature input from the encoder path. After passing through the DIPC block, it is  transformed into $224 / 2^{n-1} \times 224 / 2^{n-1} \times k$ where $k=2c$. This feeds into the next DIPC Block after a pair convolution layers. 
Here, $ c = {\{32, 32, 64,128}\}$. These feature maps belong to the layer level $n \in {1, 2, 3, 4}$ within the encoder path. At the same time, we pass the input through a max-pooling layer to generate salient feature maps. These maps possess spatial dimensions of $224 / 2^{n} \times 224 / 2^{n} \times 3$ and capture higher-level information from the downscaled versions of the feature maps.

The outputs of the involution operations are added together. This is followed by a network segment having convolution layers  with an output layer. It uses a sigmoid activation function.
This segment produces feature maps that highlight specific spatial positions, leading to improved contextual awareness amplifying the informative regions.
% This segment reduces the noise and amplifies the informative regions.

 \begin{figure*}[htbp]
  \centering
  
  \includegraphics[width=1.0\textwidth,height=8.79   cm]{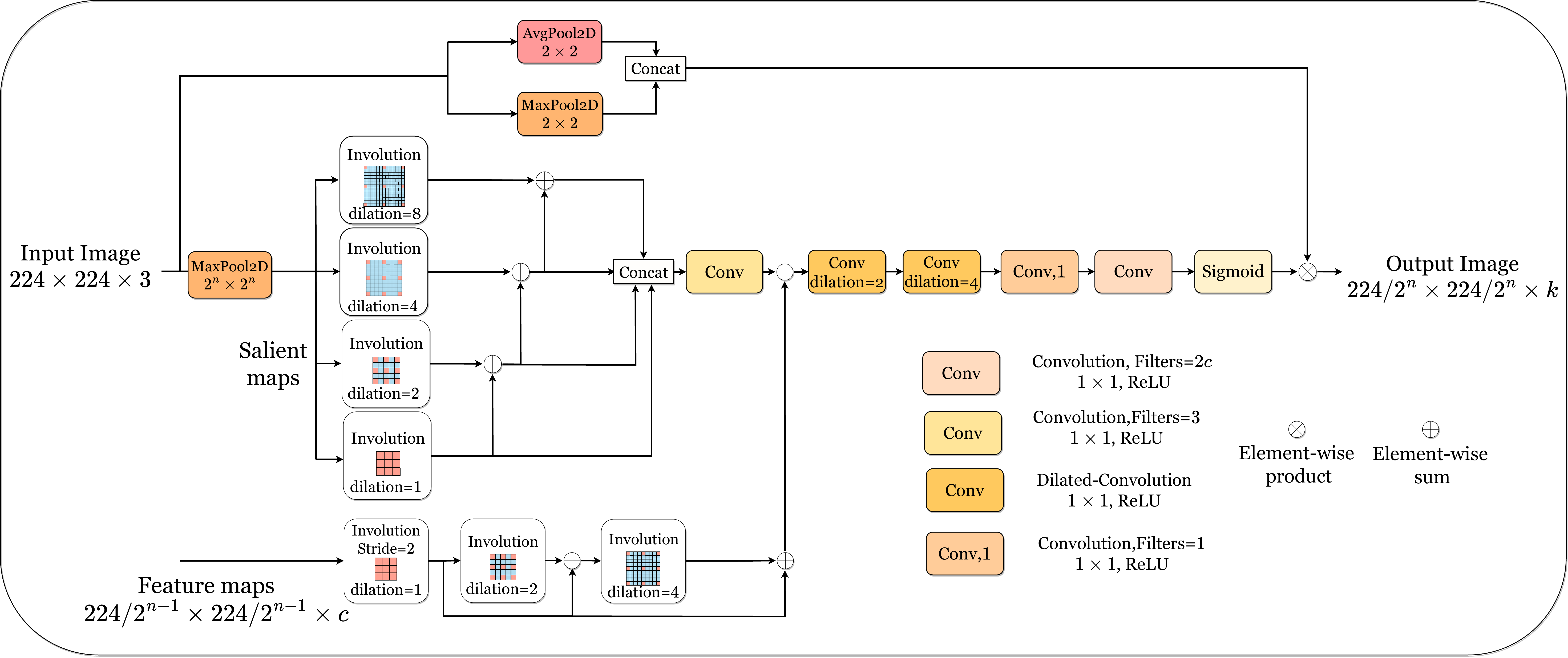}   
  \captionsetup{font=sf, justification=centering} 
  \caption{The DIPC block architecture involves a sequential transformation of input images into down-sampled maps across various stages, where $k$ \& $c$ are the number of channels with $n \in \{1,2,3,4\}$.}
  \label{fig1}
\end{figure*}

\begin{table}
\centering
\caption{Comparative results (TL=Tversky Loss)}
\label{tab:my-table}
\resizebox{\columnwidth}{!}{
\begin{tabular}{cccccc}
\hline
Class &
  Method &
  Dice &
  Jaccard &
  Precision &
  Recall \\ \hline
Normal &
  \begin{tabular}[c]{@{}c@{}}UNet\\ UNet+TL\\ FeedNet\\ AttResDUNet\\ SPEEDNet\end{tabular} &
  \begin{tabular}[c]{@{}c@{}}0.531\\ 0.945\\ 0.921\\ 0.943\\ \textbf{0.957}\end{tabular} &
  \begin{tabular}[c]{@{}c@{}}0.462\\ 0.846\\ 0.850\\ 0.773\\ \textbf{0.868}\end{tabular} &
  \begin{tabular}[c]{@{}c@{}}0.626\\ 0.923\\ 0.851\\ 0.921\\ \textbf{0.930}\end{tabular} &
  \begin{tabular}[c]{@{}c@{}}0.428\\ 0.942\\ 0.915\\ 0.927\\ \textbf{0.959}\end{tabular} \\ \hline
PolyP &
  \begin{tabular}[c]{@{}c@{}}UNet\\ UNet+TL\\ FeedNet\\ AttResDUNet\\ SPEEDNet\end{tabular} &
  \begin{tabular}[c]{@{}c@{}}0.951\\ 0.950\\ 0.952\\ 0.948\\ \textbf{0.969}\end{tabular} &
  \begin{tabular}[c]{@{}c@{}}0.301\\ 0.829\\ 0.908\\ 0.771\\ \textbf{0.877}\end{tabular} &
  \begin{tabular}[c]{@{}c@{}}0.498\\ 0.914\\ 0.865\\ 0.913\\ \textbf{0.929}\end{tabular} &
  \begin{tabular}[c]{@{}c@{}}0.471\\ 0.955\\ 0.927\\ 0.957\\ \textbf{0.972}\end{tabular} \\ \hline
High Grade-IN &
  \begin{tabular}[c]{@{}c@{}}UNet\\ UNet+TL\\ FeedNet\\ AttResDUNet\\ SPEEDNet\end{tabular} &
  \begin{tabular}[c]{@{}c@{}}0.892\\ 0.929\\ 0.848\\ 0.911\\ \textbf{0.940}\end{tabular} &
  \begin{tabular}[c]{@{}c@{}}0.810\\ 0.836\\ 0.736\\ 0.782\\ \textbf{0.864}\end{tabular} &
  \begin{tabular}[c]{@{}c@{}}0.843\\ 0.890\\ 0.896\\ 0.887\\ \textbf{0.899}\end{tabular} &
  \begin{tabular}[c]{@{}c@{}}0.960\\ 0.949\\ 0.923\\ 0.935\\ \textbf{0.978}\end{tabular} \\ \hline
Low Grade-IN &
  \begin{tabular}[c]{@{}c@{}}UNet\\ UNet+TL\\ FeedNet\\ AttResDUNet\\ SPEEDNet\end{tabular} &
  \begin{tabular}[c]{@{}c@{}}0.901\\ 0.911\\ 0.805\\ 0.946\\ \textbf{0.957}\end{tabular} &
  \begin{tabular}[c]{@{}c@{}}0.839\\ 0.860\\ 0.721\\ 0.803\\ \textbf{0.885}\end{tabular} &
  \begin{tabular}[c]{@{}c@{}}0.866\\ 0.922\\ 0.891\\ 0.916\\ \textbf{0.931}\end{tabular} &
  \begin{tabular}[c]{@{}c@{}}0.951\\ 0.963\\ 0.934\\ 0.949\\ \textbf{0.977}\end{tabular} \\ \hline
Adenocarcinoma &
   \begin{tabular}[c]{@{}c@{}}UNet\\ UNet+TL\\ FeedNet\\ AttResDUNet\\ SPEEDNet\end{tabular} &
  \begin{tabular}[c]{@{}c@{}}0.884\\ 0.897\\ 0.729\\ 0.896\\ \textbf{0.910}\end{tabular} &
  \begin{tabular}[c]{@{}c@{}}0.801\\ 0.785\\ 0.576\\ 0.744\\ \textbf{0.820}\end{tabular} &
  \begin{tabular}[c]{@{}c@{}}0.848\\ 0.862\\ 0.865\\ 0.856\\ \textbf{0.871}\end{tabular} &
  \begin{tabular}[c]{@{}c@{}}\textbf{0.950}\\ 0.928\\ 0.935\\ 0.914\\  0.948\end{tabular} \\ \hline
Serrated adenoma &
  \begin{tabular}[c]{@{}c@{}}UNet\\ UNet+TL\\ FeedNet\\ AttResDUNet\\ SPEEDNet\end{tabular} &
  \begin{tabular}[c]{@{}c@{}}0.928\\ 0.927\\ 0.883\\ 0.937\\ \textbf{0.952}\end{tabular} &
  \begin{tabular}[c]{@{}c@{}}0.881\\ 0.803\\ 0.790\\ 0.756\\ \textbf{0.899}\end{tabular} &
  \begin{tabular}[c]{@{}c@{}}0.862\\ 0.902\\ 0.887\\ 0.917\\ \textbf{0.921}\end{tabular} &
  \begin{tabular}[c]{@{}c@{}}0.980\\ 0.943\\ 0.948\\ 0.944\\ \textbf{0.986}\end{tabular} \\ \hline
Overall &
  \begin{tabular}[c]{@{}c@{}}UNet\\ UNet+TL\\ FeedNet\\ AttResDUNet\\ SPEEDNet\end{tabular} &
  \begin{tabular}[c]{@{}c@{}}0.931\\ 0.937\\ 0.823\\ 0.931\\ \textbf{0.953}\end{tabular} &
  \begin{tabular}[c]{@{}c@{}}0.843\\ 0.784\\ 0.700\\ 0.764\\ \textbf{0.865}\end{tabular} &
  \begin{tabular}[c]{@{}c@{}}0.896\\ 0.890\\ 0.781\\ 0.889\\ \textbf{0.908}\end{tabular} &
  \begin{tabular}[c]{@{}c@{}}0.961\\ 0.935\\ 0.907\\ 0.954\\ \textbf{0.971}\end{tabular} \\ \hline
  \end{tabular}
}
\end{table}
\begin{figure*}[htbp]
  \centering
   \includegraphics[width=0.9\textwidth, height=17.71cm]{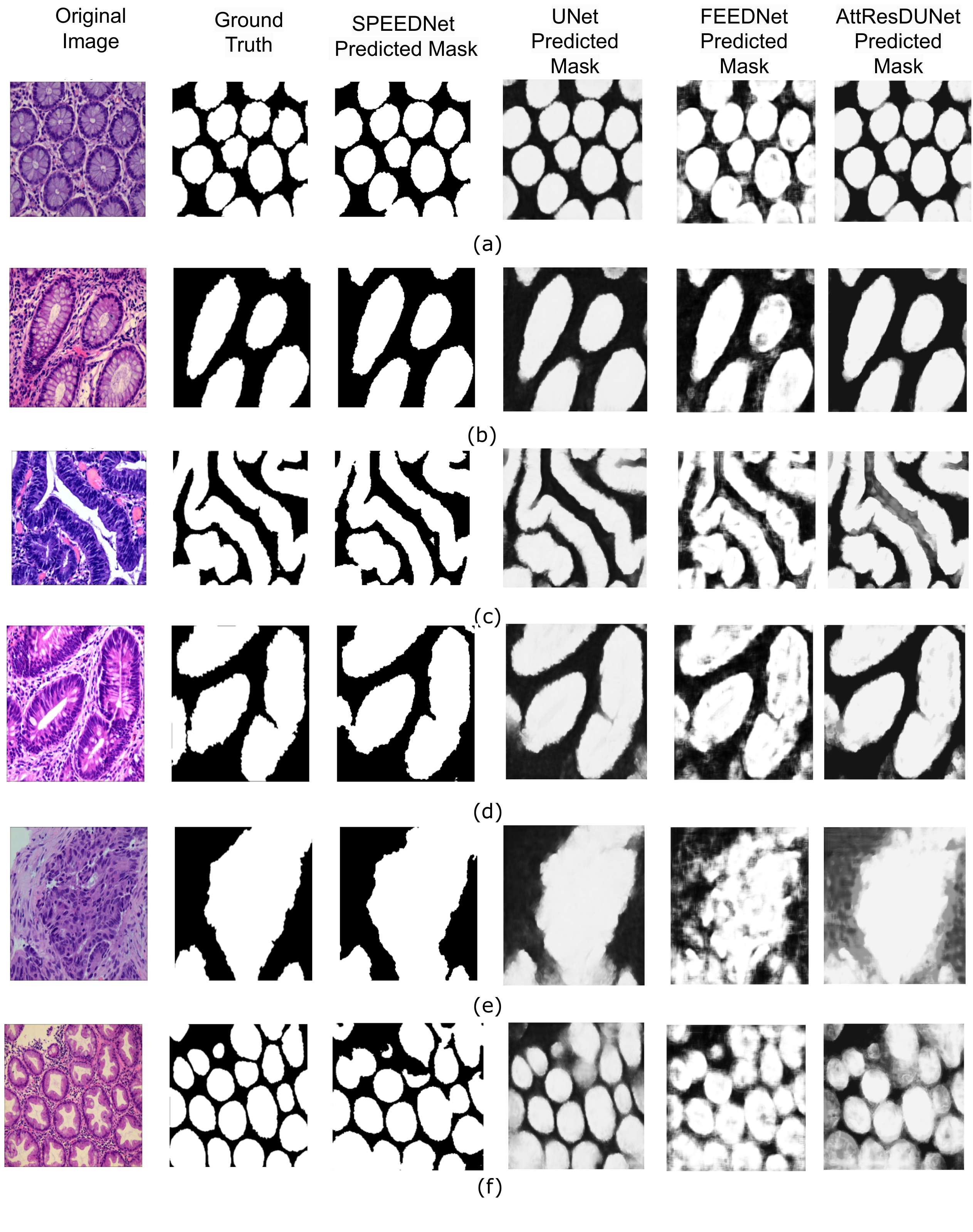}
  \captionsetup{font=sf} 
  \caption{
  Colonoscopy input images, ground truth mask, and predicted masks from SPEEDNet and UNet for the classes: (a) Normal, (b) Polyp, (c) High-grade IN, (d) Low-grade IN, (e) Adenocarcinoma, (f) Serrated adenoma.}
  \label{fig12}
\end{figure*}
\section{Experimental Setup}
\textbf{Dataset:} The EBHI-Seg dataset \cite{shi2023ebhi} comprises 6-class biopsy images from the small intestine using hematoxylin and eosin (H\&E) staining. It has 2,228 images of size 224x224, but labels are available for only 2,226 images. Hence, we discard two images that have no label. We split the dataset with a ratio of 80:20 into training and testing. For a comprehensive evaluation, we use four metrics viz., Dice Coefficient, Jaccard Index, Precision, and Recall.

\textbf{Training details:} With the Adam optimizer function, the model with 2.40M parameters is trained over 120 epochs with a batch size of 4. The initial learning rate is set to 0.001 and is decayed by a factor of 0.1 if the loss on the training dataset is not improved within 12 epochs.
To enhance segmentation in the imbalanced medical dataset, we utilize the Tversky loss function \cite{salehi2017tversky}. This function offers adaptable constants to fine-tune the penalty for distinct error types, calculated using True Positives (TP),  False Negatives (FN), and False Positives (FP). Hyperparameters $\beta$ and $\alpha$ control the emphasis on False Negatives and False Positives, respectively.

\begin{table}
\centering
\caption{Comparison of segmentation methods (UNet, Seg-Net, and MedT)}
\label{tab:first}
\resizebox{\columnwidth}{!}{
\begin{tabular}{cccccc}
\hline
Class &
  Method &
  Dice &
  Jaccard &
  Precision &
  Recall \\ \hline
Normal &
  \begin{tabular}[c]{@{}c@{}}UNet\\ Seg-Net\\ MedT\end{tabular} &
  \begin{tabular}[c]{@{}c@{}}0.531\\ 0.797\\ 0.695\end{tabular} &
  \begin{tabular}[c]{@{}c@{}}0.462\\ 0.667\\ 0.545\end{tabular} &
  \begin{tabular}[c]{@{}c@{}}0.626\\ 0.892\\ 0.867\end{tabular} &
  \begin{tabular}[c]{@{}c@{}}0.428\\ 0.743\\ 0.607\end{tabular} \\ \hline
PolyP &
  \begin{tabular}[c]{@{}c@{}}UNet\\ Seg-Net\\ MedT\end{tabular} &
  \begin{tabular}[c]{@{}c@{}}0.951\\ 0.926\\ 0.771\end{tabular} &
  \begin{tabular}[c]{@{}c@{}}0.301\\ 0.876\\ 0.634\end{tabular} &
  \begin{tabular}[c]{@{}c@{}}0.498\\ 0.890\\ 0.666\end{tabular} &
  \begin{tabular}[c]{@{}c@{}}0.471\\ 0.956\\ 0.901\end{tabular} \\ \hline
High Grade-IN &
  \begin{tabular}[c]{@{}c@{}}UNet\\ Seg-Net\\ MedT\end{tabular} &
  \begin{tabular}[c]{@{}c@{}}0.892\\ 0.886\\ 0.812\end{tabular} &
  \begin{tabular}[c]{@{}c@{}}0.810\\ 0.801\\ 0.697\end{tabular} &
  \begin{tabular}[c]{@{}c@{}}0.843\\ 0.872\\ 0.728\end{tabular} &
  \begin{tabular}[c]{@{}c@{}}0.960\\ 0.908\\ 0.945\end{tabular} \\ \hline
Low Grade-IN &
  \begin{tabular}[c]{@{}c@{}}UNet\\ Seg-Net\\ MedT\end{tabular} &
  \begin{tabular}[c]{@{}c@{}}0.901\\ 0.918\\ 0.887\end{tabular} &
  \begin{tabular}[c]{@{}c@{}}0.839\\ 0.856\\ 0.798\end{tabular} &
  \begin{tabular}[c]{@{}c@{}}0.866\\ 0.875\\ 0.866\end{tabular} &
  \begin{tabular}[c]{@{}c@{}}0.951\\ 0.967\\ 0.922\end{tabular} \\ \hline
Adenocarcinoma &
  \begin{tabular}[c]{@{}c@{}}UNet\\ Seg-Net\\ MedT\end{tabular} &
  \begin{tabular}[c]{@{}c@{}}0.884\\ 0.856\\ 0.723\end{tabular} &
  \begin{tabular}[c]{@{}c@{}}0.801\\ 0.755\\ 0.576\end{tabular} &
  \begin{tabular}[c]{@{}c@{}}0.848\\ 0.778\\ 0.645\end{tabular} &
  \begin{tabular}[c]{@{}c@{}}0.950\\ 0.977\\ 0.854\end{tabular} \\ \hline
Serrated Adenoma &
  \begin{tabular}[c]{@{}c@{}}UNet\\ Seg-Net\\ MedT\end{tabular} &
  \begin{tabular}[c]{@{}c@{}}0.928\\ 0.896\\ 0.667\end{tabular} &
  \begin{tabular}[c]{@{}c@{}}0.881\\ 0.823\\ 0.493\end{tabular} &
  \begin{tabular}[c]{@{}c@{}}0.862\\ 0.851\\ 0.876\end{tabular} &
  \begin{tabular}[c]{@{}c@{}}0.980\\ 0.923\\ 0.544\end{tabular} \\ \hline
\end{tabular}
}
\end{table}

\section{Results}
\textbf{Quantitative results:} 
We compare SPEEDNet with UNet \cite{ronneberger2015u}, AttesResDUNet \cite{khan2023attresdu}, and FeedNet \cite{deshmukh2022feednet}. We also evaluate a variant of UNet, called ``UNet+TL'', where we replace the Dice loss with the Tversky loss. 
As shown in Table \ref{tab:my-table}, SPEEDNet consistently outperforms the previous works for nearly all classes and metrics. UNet+TL outperforms UNet but still provides inferior results than SPEEDNet. In medical image segmentation, certain classes (in our case, High Grade-IN) have a very low pixel count, resulting in higher recall but worse precision. 
This shows that SPEEDNet has a high recall for positive situations. Here, positive situations refer to cases where the model correctly identifies and classifies regions of interest, i.e., true positives.

SPEEDNet achieves more than 94\% recall across all classes. Notably, for Normal and PolyP classes, SPEEDNet attains remarkable recall scores of 95.9\% and 97.2\%, and outperforms other networks by a substantial margin.
The EBHI-SEG paper  \cite{shi2023ebhi} has shown that UNet provides superior results than SegNet \cite{badrinarayanan2017segnet} and MedT \cite{valanarasu2021medical}. Since SPEEDNet outperforms UNet with tversky loss, it also outperforms traditional UNet, SegNet and MedT also as shown in Table \ref{tab:first}.

\textbf{Qualitative results:}
Fig. \ref{fig12} depicts the segmentation results for a sample image from each of the six classes. Notice that the masks produced by other netwroks have a patchy and blurred effect. The exceptional result on EBHI-Seg indicates SPEED-Net's capacity to generalize to complex data. Because of the class imbalance, the predicted mask for Serrated adenoma differs visibly from the ground truth. Including more images of this class in the dataset may improve the predicted mask quality.

\textbf{Ablation Studies:} To gain further insights, we now present ablation results (refer Table \ref{tab:ablation-table}). 

\begin{table}[htbp]
\caption{Ablation results }
\label{tab:ablation-table}
\resizebox{\columnwidth}{!}{
\begin{tabular}{cccccc}
\hline
Class &
  Method &
  Dice &
  Jaccard &
  Precision &
  Recall \\ \hline
Normal &
  \begin{tabular}[c]{@{}c@{}}No Involution\\ SPEEDNet+DB\\ SPEEDNet\end{tabular} &
  \begin{tabular}[c]{@{}c@{}}0.940\\ 0.936\\ \textbf{0.957}\end{tabular} &
  \begin{tabular}[c]{@{}c@{}}0.843\\ 0.856\\ \textbf{0.868}\end{tabular} &
  \begin{tabular}[c]{@{}c@{}}0.906\\ 0.909\\ \textbf{0.930}\end{tabular} &
  \begin{tabular}[c]{@{}c@{}}0.945\\ 0.946\\ \textbf{0.959}\end{tabular} \\ \hline
PolyP &
  \begin{tabular}[c]{@{}c@{}}No Involution\\ SPEEDNet+DB\\ SPEEDNet\end{tabular} &
  \begin{tabular}[c]{@{}c@{}}0.941\\ 0.943\\ \textbf{0.969}\end{tabular} &
  \begin{tabular}[c]{@{}c@{}}0.833\\ 0.858\\ \textbf{0.877}\end{tabular} &
  \begin{tabular}[c]{@{}c@{}}0.883\\ 0.903\\ \textbf{0.929}\end{tabular} &
  \begin{tabular}[c]{@{}c@{}}0.962\\ 0.964\\ \textbf{0.972}\end{tabular} \\ \hline
High Grade-IN &
  \begin{tabular}[c]{@{}c@{}}No Involution\\ SPEEDNet+DB\\ SPEEDNet\end{tabular} &
  \begin{tabular}[c]{@{}c@{}}0.924\\ 0.926\\ \textbf{0.940}\end{tabular} &
  \begin{tabular}[c]{@{}c@{}}0.835\\ 0.837\\ \textbf{0.864}\end{tabular} &
  \begin{tabular}[c]{@{}c@{}}0.887\\ 0.881\\ \textbf{0.896}\end{tabular} &
  \begin{tabular}[c]{@{}c@{}}0.960\\ 0.961\\ \textbf{0.978}\end{tabular} \\ \hline
Low Grade-IN &
  \begin{tabular}[c]{@{}c@{}}No Involution\\ SPEEDNet+DB\\ SPEEDNet\end{tabular} &
  \begin{tabular}[c]{@{}c@{}}0.942\\ 0.944\\ \textbf{0.957}\end{tabular} &
  \begin{tabular}[c]{@{}c@{}}0.871\\ 0.874\\ \textbf{0.885}\end{tabular} &
  \begin{tabular}[c]{@{}c@{}}0.913\\ 0.917\\ \textbf{0.931}\end{tabular} &
  \begin{tabular}[c]{@{}c@{}}0.963\\ 0.968\\ \textbf{0.977}\end{tabular} \\ \hline
Adenocarcinoma &
  \begin{tabular}[c]{@{}c@{}}No Involution\\ SPEEDNet+DB\\ SPEEDNet\end{tabular} &
  \begin{tabular}[c]{@{}c@{}}0.895\\ 0.892\\ \textbf{0.910}\end{tabular} &
  \begin{tabular}[c]{@{}c@{}}0.794\\ 0.809\\ \textbf{0.820}\end{tabular} &
  \begin{tabular}[c]{@{}c@{}}0.850\\ 0.858\\ \textbf{0.871}\end{tabular} &
  \begin{tabular}[c]{@{}c@{}}0.931\\ 0.937\\ \textbf{0.946}\end{tabular} \\ \hline
Serrated adenoma &
  \begin{tabular}[c]{@{}c@{}}No Involution\\ SPEEDNet+DB\\ SPEEDNet\end{tabular} &
  \begin{tabular}[c]{@{}c@{}}0.911\\ 0.930\\ \textbf{0.952}\end{tabular} &
  \begin{tabular}[c]{@{}c@{}}0.842\\ 0.840\\ \textbf{0.899}\end{tabular} &
  \begin{tabular}[c]{@{}c@{}}0.879\\ 0.897\\ \textbf{0.921}\end{tabular} &
  \begin{tabular}[c]{@{}c@{}}0.975\\ 0.951\\ \textbf{0.983}\end{tabular} \\ \hline
\end{tabular}
}
\end{table}
\textit{1. No Involution:} Replacing the involution with convolution not only degrades all metrics but also increases the parameters from 2.40 million to 4.95 million.

\textit{2. SPEEDNet with Dilated Bottleneck:} We add a dilation rate to the convolution layers in the bottleneck. This provides better results than not using involution, but remains inferior to our full network.

\section{Conclusion}
This study proposed SPEEDNet, a lightweight network developed by utilizing a novel DIPyramidal Convolution Fusion Block that minimizes computational complexity via involution. The approach represents a novel direction in network architecture, focusing on effective segmentation while minimizing resource demands. Our forthcoming endeavors will center around fusing additional contextual information, such as temporal sequences or multi-modal data and using unsupervised learning to improve segmentation performance.

Overall, our proposed SPEEDNet architecture offers a promising solution for medical image segmentation problems with the potential to substantially enhance diagnostic accuracy and efficiency.

% \begin{figure}
%     \centering
%     \includegraphics[width=1.01\columnwidth]{abl.pdf}
%     \caption{Metrics for ablation study.}
%     \label{fig:enter-label}
% \end{figure}

\bibliographystyle{IEEEtran}
\bibliography{ref.bib}

\begin{thebibliography}{10}
\providecommand{\url}[1]{#1}
\csname url@samestyle\endcsname
\providecommand{\newblock}{\relax}
\providecommand{\bibinfo}[2]{#2}
\providecommand{\BIBentrySTDinterwordspacing}{\spaceskip=0pt\relax}
\providecommand{\BIBentryALTinterwordstretchfactor}{4}
\providecommand{\BIBentryALTinterwordspacing}{\spaceskip=\fontdimen2\font plus
\BIBentryALTinterwordstretchfactor\fontdimen3\font minus
  \fontdimen4\font\relax}
\providecommand{\BIBforeignlanguage}[2]{{%
\expandafter\ifx\csname l@#1\endcsname\relax
\typeout{** WARNING: IEEEtran.bst: No hyphenation pattern has been}%
\typeout{** loaded for the language `#1'. Using the pattern for}%
\typeout{** the default language instead.}%
\else
\language=\csname l@#1\endcsname
\fi
#2}}
\providecommand{\BIBdecl}{\relax}
\BIBdecl

\bibitem{litjens2016deep}
G.~Litjens, C.~I. S{\'a}nchez, N.~Timofeeva, M.~Hermsen, I.~Nagtegaal,
  I.~Kovacs, C.~Hulsbergen-Van De~Kaa, P.~Bult, B.~Van~Ginneken, and J.~Van
  Der~Laak, ``Deep learning as a tool for increased accuracy and efficiency of
  histopathological diagnosis,'' \emph{Scientific reports}, vol.~6, no.~1, p.
  26286, 2016.

\bibitem{tajbakhsh2015automated}
N.~Tajbakhsh \emph{et~al.}, ``Automated polyp detection in colonoscopy videos
  using shape and context information,'' \emph{IEEE TMI}, 2015.

\bibitem{badrinarayanan2017segnet}
V.~Badrinarayanan \emph{et~al.}, ``Segnet: A deep convolutional encoder-decoder
  architecture for image segmentation,'' \emph{TPAMI}, 2017.

\bibitem{jha2019resunet++}
D.~Jha \emph{et~al.}, ``Resunet++: An advanced architecture for medical image
  segmentation,'' in \emph{IEEE ISM}, 2019, pp. 225--2255.

\bibitem{graham2019mild}
S.~Graham \emph{et~al.}, ``{MILD-Net: Minimal information loss dilated network
  for gland instance segmentation in colon histology images},'' \emph{Medical
  image analysis}, vol.~52, pp. 199--211, 2019.

\bibitem{dumitru2023using}
R.-G. Dumitru, D.~Peteleaza, and C.~Craciun, ``Using duck-net for polyp image
  segmentation,'' \emph{Scientific Reports}, vol.~13, no.~1, p. 9803, 2023.

\bibitem{shi2023ebhi}
L.~Shi, X.~Li, W.~Hu, H.~Chen, J.~Chen, Z.~Fan, M.~Gao, Y.~Jing, G.~Lu, D.~Ma
  \emph{et~al.}, ``Ebhi-seg: A novel enteroscope biopsy histopathological
  hematoxylin and eosin image dataset for image segmentation tasks,''
  \emph{Frontiers in Medicine}, vol.~10, p. 1114673, 2023.

\bibitem{ronneberger2015u}
O.~Ronneberger, P.~Fischer, and T.~Brox, ``U-net: Convolutional networks for
  biomedical image segmentation,'' in \emph{MICCAI}, 2015, pp. 234--241.

\bibitem{deshmukh2022feednet}
G.~Deshmukh \emph{et~al.}, ``Feednet: A feature enhanced encoder-decoder lstm
  network for nuclei instance segmentation for histopathological diagnosis,''
  \emph{Physics in Medicine \& Biology}, 2022.

\bibitem{vakanski2020attention}
A.~Vakanski, M.~Xian, and P.~E. Freer, ``Attention-enriched deep learning model
  for breast tumor segmentation in ultrasound images,'' \emph{Ultrasound in
  medicine \& biology}, vol.~46, no.~10, pp. 2819--2833, 2020.

\bibitem{salehi2017tversky}
S.~S.~M. Salehi \emph{et~al.}, ``Tversky loss function for image segmentation
  using 3d fully convolutional deep networks,'' in \emph{MLMI}, 2017.

\bibitem{khan2023attresdu}
A.~M. Khan \emph{et~al.}, ``Attresdu-net: Medical image segmentation using
  attention-based residual double u-net,'' \emph{arXiv preprint
  arXiv:2306.14255}, 2023.

\bibitem{valanarasu2021medical}
J.~M.~J. Valanarasu \emph{et~al.}, ``Medical transformer: Gated axial-attention
  for medical image segmentation,'' in \emph{MICCAI}, 2021, pp. 36--46.

\end{thebibliography}
\end{document}